\documentclass[conference,10pt]{IEEEtran}
\usepackage{amsmath,graphicx,cite,color}
\usepackage[ruled,linesnumbered]{algorithm2e}
\usepackage{amsfonts,amssymb}
\usepackage{multirow}
\usepackage{balance}

\IEEEoverridecommandlockouts

\newcommand{\A}{{\bf A}}

\newcommand{\x}{{\bf x}}

\newcommand{\I}{{\bf I}}

\begin{document}

\title{
Independent Vector Extraction Constrained on Manifold of Half-Length Filters}

\author{\authorblockN{Zbyn\v{e}k Koldovsk\'{y}\authorrefmark{1}, Jaroslav \v{C}mejla\authorrefmark{1}, T\"ulay Adali\authorrefmark{2}, and Stephen O'Regan\authorrefmark{3}
}
\authorblockA{\authorrefmark{1}
Acoustic Signal Analysis and Processing Group, Faculty of Mechatronics, Informatics, 
and Interdisciplinary Studies,\\ Technical University of Liberec, Czech Republic.}
\authorblockA{\authorrefmark{2}Department of Electrical and Computer Engineering\\
  University of Maryland, Baltimore County, Baltimore, MD, 21250, USA}
\authorblockA{\authorrefmark{3}Naval Surface Warfare Center Carderock Division, West Bethesda, Maryland, USA}
\thanks{This work was supported by the department of the Navy, Office of Naval Research Global, through Project No.~N62909-19-1-2105.} 
}

\maketitle
\begin{abstract}
Independent Vector Analysis (IVA) is a popular extension of Independent Component Analysis (ICA) for joint separation of a set of instantaneous linear mixtures, with a direct application in frequency-domain speaker separation or extraction. The mixtures are parameterized by mixing matrices, one matrix per mixture. This means that the IVA mixing model does not account for any relationships between parameters across the mixtures/frequencies. The separation proceeds jointly only through the source model, where statistical dependencies of sources across the mixtures are taken into account. 
In this paper, we propose a mixing model for joint blind source extraction where the mixing model parameters are linked across the frequencies. This is achieved by constraining the set of feasible parameters to the manifold of half-length separating filters, which has a clear interpretation and application in frequency-domain speaker extraction. 
\end{abstract}
\begin{keywords}
	Blind Source Separation, Blind Source Extraction, Independent Component Analysis, Independent Vector Analysis, Speaker Extraction
\end{keywords}
\section{Introduction}
\label{sec:intro}
Independent Component and Vector Analysis (ICA/IVA) are popular methods for Blind Source Separation (BSS) based on the assumption that observed signal mixtures consist of independent original signals (sources) \cite{comon2010handbook}. In ICA, only one mixture of independent signals is considered; when there are $K>1$ mixtures, each mixture is processed independently. In IVA, $K>1$ mixtures are treated jointly by exploiting dependencies among sources from different mixtures \cite{kim2007}. Independent Component/Vector Extraction (ICE/IVE) are related methods to ICA/IVA for Blind Source Extraction (BSE) where the goal is to extract only a particular source-of-interest (SOI) from every mixture, i.e., to separate it from the other signals \cite{koldovsky2019TSP,scheibler2019overiva,ikeshita2021independent}. 

Inherent in the BSS and BSE tasks are uncertainties, which in practice correspond to crucial problems such as determining which speaker is the target in the speaker extraction problem. The single-mixture problem can be solved up to the order and scaling of the original signals. When considering $K>1$ mixtures, the unknown order causes the permutation problem, which means that the separated sources have different orders in each mixture \cite{sawada2006sap}. It is one of the ideas behind the IVA that if mixtures are separated jointly, and dependencies between sources across the mixtures are exploited, this will help to solve the permutation problem \cite{kim2006,lee2007fast}.

Specifically, in IVA, we consider $K$ linear instantaneous mixtures 
\begin{equation}\label{eq:modelstaticICA}
\x^k = \A^k {\bf s}^k,\qquad k=1,\dots,K,
\end{equation} 
where $\x^k$ is a $d\times 1$ vector of random variables (RVs) representing the mixed (observed) signals; $\A^k$ is a $d\times d$ non-singular mixing matrix; and ${\bf s}^k$ is a $d\times 1$ vector of independent RVs representing the original sources; $k$ is the mixture index. Samples of signals are assumed to be independent realizations of the corresponding random variables (independently and identically distributed - i.i.d.). The goal is to estimate de-mixing matrices ${\bf W}^k$ such that ${\bf y}^k={\bf W}^k\x^k$ correspond to ${\bf s}^k$ up to the indeterminable scales and order. ${\bf W}^k$ are sought such that the separated signals ${\bf y}^k$ are independent. The $i$th sources in ${\bf s}^1,\dots,{\bf s}^K$, represented by the elements of vector random variables ${\bf s}_i=[s^1_i,\dots,s^K_i]^T$, $i=1,\dots,d$, are modeled as dependent through joint distributions, which helps in solving the permutation problem\cite{kim2006}. 

In speaker extraction, IVA or IVE is deployed in the frequency domain as follows: The mixed signals observed through microphones in the time domain are transformed by the short-term Fourier transform (STFT) \cite{ASSSEbook2018}. The model \eqref{eq:modelstaticICA} then describes the observed signals in the $k$th frequency band whose samples correspond to frames. The mixing matrices ${\bf A}^k$ represent acoustic transfer functions between the speakers and microphones corresponding to room impulse responses, and the de-mixing matrices ${\bf W}^k$ represent transfer functions of separating filters. The permutation problem must be solved to align the separated frequency components of each speaker. To do this, IVA uses dependencies between $s^1_i,\dots,s^K_i$, which can be interpreted as dependencies between frequency components of the same speaker \cite{kim2007}. 

While IVA and IVE are successful, there are, at least, two doubtful aspects. Looking back at \eqref{eq:modelstaticICA}, one can see that the model of dependencies can be fragile since it purely relies on higher-order statistical relationships \cite{kim2007}. It is known that the solution to the permutation problem by IVA/IVE is not guaranteed and many imperfections appear; \cite{koldovsky2022double}. Second, the model's parameterization increases rapidly with the frequency resolution of the STFT. It is clear that from a certain $K$, the separation accuracy will no longer improve while the computational demands and difficulty of the problem will increase \cite{araki2003sap}. 


In this paper, we propose a modified algorithm for IVE that operates on a constrained set of feasible (de-)mixing matrices in \eqref{eq:modelstaticICA}. Particularly, we consider the special case where the set corresponds to the manifold of half-length separating filters. Our approach gives guidance for deriving other similar modifications of IVE (and IVA) and makes several senses both in terms of frequency-domain speaker extraction as well as in terms of IVE and IVA as methods for general joint BSE/BSS. The experiments show some advantageous properties of the proposed (gradient) algorithm in terms of computational savings, convergence speed, and extraction accuracy.


The paper is organized as follows. The problem is formulated in Section~II and algorithm is derived in Section~III. Section~IV presents two experimental evaluations, and Section~V concludes this paper. 

{\em Notations:} 
Plain, bold, and bold capital letters denote, respectively, scalars, vectors, and matrices. Upper index $\cdot^T$, $\cdot^H$, or 
$\cdot^*$ denotes, respectively, transposition, conjugate transpose, or complex 
conjugate. The Matlab convention for matrix/vector concatenation will be used, e.g., $[1;\,{\bf g}]=[1,\, {\bf g}^T]^T$. 
${\rm E}[\cdot]$ stands for the expectation operator, and $\hat{\rm E}[\cdot]$ is the sample-based average taken over all available samples of the argument. 

\section{Problem Formulation}
\subsection{Independent Vector Extraction}
We first describe the modification of the mixing model \eqref{eq:modelstaticICA} for the joint extraction problem; for more details see \cite{koldovsky2019TSP}. Without loss of generality, let the SOI, in every mixture, be the first source signal $s^k_1$, $k=1,\dots,K$. We can rewrite \eqref{eq:modelstaticICA} as
\begin{equation}\label{mixing_determined}
    {\bf x}^k = {\bf A}^k{\bf u}^k,
\end{equation}
where $\A^k=[{\bf a}^k,\,{\bf Q}^k]$ and ${\bf u}^k=[s^k_1;\,{\bf z}^k]$. When only $s^k_1$ should be extracted from ${\bf x}^k$, neither ${\bf Q}$ nor ${\bf z}[n]$ need to be identified and separated, only their corresponding subspaces. These actually correspond to the other ``background'' signals in ${\bf x}^k$. In \cite{koldovsky2019TSP}, it is shown that, for this task, the mixing matrix ${\bf A}^k$ and the de-mixing matrix ${\bf W}^k=({\bf A}^k)^{-1}$ can be parameterized by the mixing vector ${\bf a}^k$, which is the first column of ${\bf A}^k$ in \eqref{eq:modelstaticICA},  and by the {\em separating vector} $({\bf w}^k)^H$, which is the first row of ${\bf W}$, as follows:
\begin{align}\label{BSE_mixing_matrix}
    {\bf A}^k&= 
 \begin{pmatrix}
  \gamma^k & ({\bf h}^k)^H\\
   {\bf g}^k &  \frac{1}{\gamma^k}({\bf g}^k({\bf h}^k)^H-{\bf I}_{d-1})
    \end{pmatrix},\\
   {\bf W}^k &=
   \begin{pmatrix}
   ({\bf w}^k)^H\\
   {\bf B}^k
   \end{pmatrix}=
     \begin{pmatrix}
     (\beta^k)^* & ({\bf h}^k)^H\\
     {\bf g}^k & -\gamma^k {\bf I}_{d-1}
     \end{pmatrix},
\end{align}
where ${\bf a}^k=[\gamma^k;\,{\bf g}^k]$ and ${\bf w}=[\beta;\,{\bf h}^k]$; the parameter vectors ${\bf a}^k$ and ${\bf w}^k$ are linked through the condition $({\bf a}^k)^H{\bf w}_{k}=1$; $\I_{d}$ stands for the $d\times d$ identity matrix. The extracted signal is obtained as $s^k_1=({\bf w}^k)^H{\bf x}^k$ and the background signals are ${\bf z}^k={\bf B}^k{\bf x}^k$.
To simplify the notation, we further denote the vector component corresponding to the SOIs by ${\bf s}=[s^1,\dots,s^K]^T$ where $s^k=s^k_1$, $k=1,\dots,K$.

\subsection{Source model, contrast function, and gradient algorithm}
The basic source model used in IVE assumes that ${\bf s}$ is distributed according to a joint non-Gaussian pdf $p({\bf s})$, and ${\bf z}^k$, $k=1,\dots,K$, are circular Gaussian; ${\bf z}^{k_1}$ and ${\bf z}^{k_2}$ are assumed to be uncorrelated for $k_1\neq k_2$ \cite{koldovsky2019TSP}. Most importantly, ${\bf s}$ and ${\bf z}^k$ are independent. Since $p({\bf s})$ is unknown, it must be replaced by a suitable model density. In \cite{koldovsky2021fastdiva} (Section~III.B), it is proposed to replace $p({\bf s})$ by
\begin{equation}\label{eq:modeldensity}
    p({\bf s}) \approx f\left(\overline{ \bf s}\right)\left(\prod_{k=1}^K\sigma_k\right)^{-2},
\end{equation}
where $f(\cdot)$ is a suitable normalized non-Gaussian pdf, and $\sigma_{k}^2$ is the sample-based variance of the estimate of $s^{k}$, and $\overline{\bf s}$ consists of normalized elements of ${\bf s}$. It is worth noting that, in most ICA/IVA-related methods, $\sigma_{k}^2$ are assumed to be equal to $1$, which is a way to cope with the scaling ambiguity of $s^k$. Later in this paper, we will show that a different scaling ambiguity treatment is needed in our case.

The contrast function for estimating the model parameters is given by \cite{koldovsky2021fastdiva}
\begin{multline}\label{eq:contastIVE}
    \mathcal{C}\left(\{{\bf w}^k,{\bf a}^k\}_{k=1,\dots,K}\right) =\hat{\rm E}\left[\log f\left(\overline{\bf s}\right)\right]  -\sum_{k=1}^{K}\log\sigma_{k}^2\\ -\sum_{k=1}^{K} \hat{\rm E}\left[({\bf z}^{k})^H({\bf C}_{\bf z}^{k})^{-1}{\bf z}^{k}\right]  
    + (d-2)\sum_{k=1}^{K}\log |\gamma^{k}|^2 + C,
\end{multline}
where ${\bf C}_{\bf z}^{k}$ denotes the covariance matrix of ${\bf z}^{k}$; $C$ is a constant independent of the parameter vectors. The function is derived from the likelihood function in which the unknown SOI pdf is replaced by \eqref{eq:modeldensity}, and ${\bf C}_{\bf z}^{k}$ is replaced by the current sample-based estimate \cite{koldovsky2021fastdiva}.

The estimates of ${\bf a}^k$ and ${\bf w}^k$ are sought through maximizing \eqref{eq:contastIVE} under the orthogonal constraint that enforces that the current estimates of $s^k$ and ${\bf z}^k$ have sample correlation equal to zero. It makes ${\bf a}^{k}$ fully dependent on ${\bf w}^k$ through ${\bf a}^k={\bf C}^k{\bf w}^k/\sigma^2_k$, where ${\bf C}^k$ is the covariance matrix of ${\bf x}^k$. After a normalization step due to the replacement of the true pdf by \eqref{eq:modeldensity}, the gradient of \eqref{eq:contastIVE} by $({\bf w}^k)^H$ under the orthogonal constraint reads 
\begin{equation}
\frac{\partial\mathcal{C}}{\partial({\bf w}^k)^*}
=\nabla^k= {\bf a}^k-\nu_k^{-1}\hat{\rm E}\left[\phi_k\left(\overline{\bf s}\right) \overline{\bf x}^k\right],\label{eq:gradient}
\end{equation}
where $\overline{\bf x}^k={\bf x}^k/\sigma_k$, $\nu_k=\hat{\rm E}[\phi_k\left(\overline{\bf s}\right)\overline{s}^k]$ is the normalization factor of the model pdf, and $\phi_k(\cdot)=-\frac{\partial}{\partial s_k}\log f(\cdot)$ is the $k$th score function of the model pdf $f(\cdot)$; see \cite{koldovsky2021fastdiva} for detailed computations.

Gradient-based algorithms seek the maximum of \eqref{eq:contastIVE} by updating ${\bf w}^k$ in the direction of the gradient, i.e. 
\begin{equation}\label{eq:iteration}
    {\bf w}^k\leftarrow {\bf w}^k+\mu\nabla^k,\quad k=1,\dots,K,
\end{equation}
where $\mu$ is a step-size parameter. In \cite{koldovsky2019TSP}, such an algorithm is referred to as OGIVE$_{\bf w}$. The important fact that follows from \eqref{eq:gradient} and \eqref{eq:iteration} is that the updates of ${\bf w}^1,\dots,{\bf w}^K$ proceed almost independently. The mutual influence of these ``parallel'' algorithms takes place through the score functions $\phi_k$, $k=1,\dots,K$, which all depend on the vector component ${\bf s}$. This decoupling is the main advantage of why most blind speech separation and extraction methods operate in the STFT domain \cite{smaragdis1998,ASSSEbook2018}. 

\subsection{Linking the parameters across frequencies}\label{sec:general}
Let us introduce a matrix
\begin{equation}
{\bf W}=[{\bf w}^1,\dots,{\bf w}^K],    
\end{equation}
whose columns correspond to the separating vectors from all frequencies. So far, the elements of ${\bf W}$ are the free variables subject to which the contrast function \eqref{eq:contastIVE} is optimized since the other parameters are dependent through the orthogonal constraint. In general, our goal is to define a new set of parameters given by the elements of a $d\times L$ matrix
\begin{equation}
{\bf V}=[{\bf v}^1,\dots,{\bf v}^L],    
\end{equation}
where ${\bf v}^\ell$ denotes its $\ell$th column, and $v_{\ell,i}$ denotes the $i$th element of ${\bf v}^\ell$, i.e., the $i\ell$th element of ${\bf V}$. In addition, we assume that a smooth mapping 
\begin{equation}\label{eq:mapping}
   {\bf W} = \mathcal{F}({\bf V}) 
\end{equation}
exists and is known. We want $L<K$ because such a mapping reduces the number of free parameters in the model.

Once such a mapping is given, we can apply the chain rule to compute the derivatives of \eqref{eq:contastIVE} by ${\bf V}$ and derive algorithms that perform the optimization within the new manifold. By \eqref{eq:gradient} and using the complex-valued chain rule \cite{petersen2008}, 
\begin{multline}\label{eq:chainrule}
    \frac{\partial\mathcal{C}}{\partial v_{\ell,i}^*}=
    {\tt tr}\left[
    \left(\frac{\partial\mathcal{C}}{\partial {\bf W}}\right)^T
    \frac{\partial {\bf W}}{\partial v_{\ell,i}^*}
    \right]+
    {\tt tr}\left[
    \left(\frac{\partial\mathcal{C}}{\partial {\bf W}^*}\right)^T
    \frac{\partial {\bf W}^*}{\partial v_{\ell,i}^*}
    \right]\\
    ={\tt tr}\left[
    \boldsymbol\Omega^H
    \frac{\partial \mathcal{F}({\bf V})}{\partial v_{\ell,i}^*}
    \right]+
    {\tt tr}\left[
    \boldsymbol\Omega^T
    \frac{\partial \mathcal{F}({\bf V})^*}{\partial v_{\ell,i}^*}
    \right],
\end{multline}
where ${\tt tr}[\cdot]$ denotes the trace of the matrix argument, and 
\begin{equation}
\boldsymbol\Omega=[\nabla^1,\dots,\nabla^K]
\end{equation}
is a matrix collecting the gradients given by \eqref{eq:gradient}.
A new gradient algorithm can be obtained when \eqref{eq:iteration} is replaced by two steps
\begin{align}
   {\bf v}^\ell &\leftarrow {\bf v}^\ell + \mu\frac{\partial\mathcal{C}}{\partial {\bf v}^\ell}, \quad \ell=1,\dots,L,\\
   {\bf W} &\leftarrow \mathcal{F}({\bf V}).
\end{align}
The increase in computational complexity depends mainly on the form of the derivatives of the mapping $\mathcal{F}(\cdot)$ in \eqref{eq:chainrule}.

\section{Proposed Algorithm}
\subsection{Half-length Filter Manifold}
We now present a variant of the general approach proposed in Section~\ref{sec:general}, tailored to the speaker extraction approach in the STFT domain. Here, the rows of ${\bf W}$ correspond to the representation of time-domain separating filters in the Discrete Fourier Transform (DFT) domain of resolution $K$. The idea is to constrain ${\bf W}$ to the manifold of filters whose length is only $K/2$. For simplicity, we will assume that $K/4$ is an integer.

To find the mapping \eqref{eq:mapping}, let us consider an FIR filter $w(n)$, $n=0,\dots,K-1$, whose length is $K/2$, i.e., $w(n)=0$ for $n=K/2,\dots,K-1$. The $k$th coefficient of its transfer function in the DFT domain of resolution $K$ is given by
\begin{equation}
    \tilde{w}_{K}(k)=\sum_{n=0}^{K-1}w(n)\mathsf{W}^{kn}_K
    =\sum_{n=0}^{\frac{K}{2}-1}w(n)\mathsf{W}^{kn}_K,
\end{equation}
where $\mathsf{W}_{K}=e^{-\tfrac{i2\pi}{K}}$. Now we express $\tilde{w}_{K}(k)$ separately for odd and even $k$.
\begin{align}
    \tilde{w}_{K}(2\ell)&=\sum_{n=0}^{\frac{K}{2}-1}w(n)\mathsf{W}^{2\ell n}_K=\sum_{n=0}^{\frac{K}{2}-1}w(n)\mathsf{W}^{\ell n}_{\frac{K}{2}}\nonumber\\
    &=
    \tilde{w}_{K/2}(\ell),\label{even}\\
    \tilde{w}_{K}(2\ell+1)&=\sum_{n=0}^{\frac{K}{2}-1}w(n)\mathsf{W}^{(2\ell+1)n}_K=\sum_{n=0}^{\frac{K}{2}-1}w(n)\mathsf{W}_K^n\mathsf{W}^{\ell n}_{\frac{K}{2}}\nonumber\\
    &=\{\tilde{w}_{K/2} \circledast \upsilon\}(\ell),\label{odd}
\end{align}
where $\ell=0,\dots,K/2-1$, $\upsilon(k)$ denotes the DFT (of length $K/2$) of the sequence $\mathsf{W}_K^n$, $n=0,\dots,K/2-1$, and $\circledast$ stands for the circular convolution. The latter relation in \eqref{odd} follows from the well-known properties of the DFT \cite{porat1997}. 

The equations \eqref{even}-\eqref{odd} describe the mapping between $\tilde{w}_{K}$ and $\tilde{w}_{K/2}$ for every FIR filter $w(n)$ of maximum length $K/2$. Since we want to constrain ${\bf W}$ to such filters, we can parameterize ${\bf W}$ through \eqref{eq:mapping} when ${\bf V}$ is such that ${\bf v}^\ell={\bf w}^{2\ell-1}$ for $\ell=1,\dots,L$, and $L=K/2$, because of \eqref{even}. The mapping is linear and is represented by an $L\times 2L$ matrix ${\bf F}$ such that\footnote{Note the conjugate value of ${\bf F}$ in \eqref{eq:linearmapping}, which is due to the fact that the columns of ${\bf W}$, i.e. the separating vectors ${\bf w}^k$, act on the observed signals conjugated, i.e., as $s^k=({\bf w}^k)^H{\bf x}^k$.} 
\begin{equation}\label{eq:linearmapping}
    {\bf W}={\bf V}{\bf F}^*.
\end{equation} 
By \eqref{even}, the odd columns of ${\bf F}$ form the $L\times L$ identity matrix. According to \eqref{odd}, the even columns of ${\bf F}$ form a circular matrix whose first row contains the values of $\upsilon(k)$, $k=0,\dots,K/2$.

\subsection{Accounting for Conjugate Symmetry}
In speech extraction, we are actually processing real-valued signals by real-valued filters. There is thus the conjugate symmetry in the STFT (DFT) domain, and only the frequency bins $k=1,\dots,K/2+1$ should be processed, where the first and last bins are real-valued while the others are complex-valued. This must be taken into account to make the new parameterization as effective as possible.

The original parameter matrix ${\bf W}$ can thus be replaced by a truncated matrix ${\bf W}_t=[{\bf w}^1,\dots,{\bf w}^{K/2+1}]$, and, similarly, ${\bf V}$ can be replaced by ${\bf V}_t=[{\bf v}^1,\dots,{\bf v}^{L/2+1}]$. The extraction algorithm will only process data of the frequency bins $k=1,\dots,K/2+1$. 

Now, the counterpart of \eqref{eq:linearmapping} is given by
\begin{equation}\label{eq:truncmapping}
    {\bf W}_t={\bf V}_t{\bf F}_1^* + {\bf V}_t^*{\bf J}{\bf F}_2^*, 
\end{equation}
where ${\bf F}_1$ and ${\bf F}_2$ are parts of ${\bf F}$ containing, respectively, its rows $1,\dots,L/2+1$ and $L/2+2,\dots,L$; ${\bf J}$ stands for the anti-diagonal identity matrix without its first and last column. 
By using \eqref{eq:truncmapping} in \eqref{eq:chainrule} and after some manipulations, it can be shown that 
\begin{equation}\label{eq:keyrelation}
    [\nabla^1_{\bf v},\dots,\nabla^{L/2+1}_{\bf v}]=\boldsymbol{\Omega}_1{\bf F}_1^T + \boldsymbol{\Omega}_1^*{\bf F}_2^H,
\end{equation}
where 
\begin{equation}\label{eq:gradientv}
   \nabla^\ell_{\bf v} =\frac{\partial\mathcal{C}}{\partial({\bf v}^\ell)^*}, \quad \ell=1,\dots,L/2+1,
\end{equation}
and $\boldsymbol{\Omega}_1=[\nabla^1,\dots,\nabla^{K/2+1}]$.

\subsection{Coping with Scaling Ambiguity}
Owing to the scaling ambiguity, the extracted signals $s^k=({\bf w}^k)^H{\bf x}^k$ can have arbitrary scales. This freedom can cause convergence issues. Therefore, it is typical to keep ${\bf w}^k$ scaled so that $\sigma_k^2$, the variance of $s^k$, is one. However, we cannot do this in our method because rescaling each ${\bf w}^k$, $k=1,\dots,K$, generally causes them not to correspond to half-length filters. A rescaling can be applied only to the new parametric vectors ${\bf v}^\ell$, $\ell=1,\dots,L/2+1$; our choice is to normalize these vectors after every iteration. It should be noted that the scale-invariant pdf model given by \eqref{eq:modeldensity} plays an essential role in this respect. Rescaling the vectors ${\bf v}^\ell$ so that $\sigma_k^2=1$ for odd $k$ cannot guarantee that $\sigma_k^2=1$ also for even $k$. With the scale-invariant pdf model, this problem no longer arises.

\section{Experiments}

\subsection{Speech extraction}
Using room impulse response generator \cite{allen1979} implemented by E.~Habets\footnote{ https://www.audiolabs-erlangen.de/fau/professor/habets/software/rir-generator}, we simulate a situation with two speakers recorded by two microphones, whose distance to each other is $20$cm. The room is $3\times 4$m and $3$m high. The microphones are placed at positions $[1.5\,2\,1]$m, and the speakers are at positions $[1\,3\,1]$m (man) and $[2\,3\,1]$m (woman), respectively. The reverberation time $T_{60}$ is set to $100$ms and $200$ms; the sampling frequency is $16$~kHz. A mixture of the speakers of $2.5$s is generated and transformed into the STFT domain with a DFT length of $512$ samples and a shift of $128$ samples. 

The blind extraction of the female speaker is performed and is evaluated in terms of the Signal-to-Interference Ratio (SIR). The proposed algorithm is compared with the gradient OGIVE$_{\bf w}$ algorithm from \cite{koldovsky2019TSP}. OGIVE$_{\bf w}$ is considered in two variants: with and without data whitening. With whitening (denoted by ``wh.''), the algorithm involves matrix inversions, while the proposed algorithm and  OGIVE$_{\bf w}$ without whitening are free of the matrix inverse operation. The step-size parameter $\mu$ is set to the same value $0.05$ in all methods.

\begin{figure}[ht]
    \centering
    \includegraphics[width=0.49\linewidth]{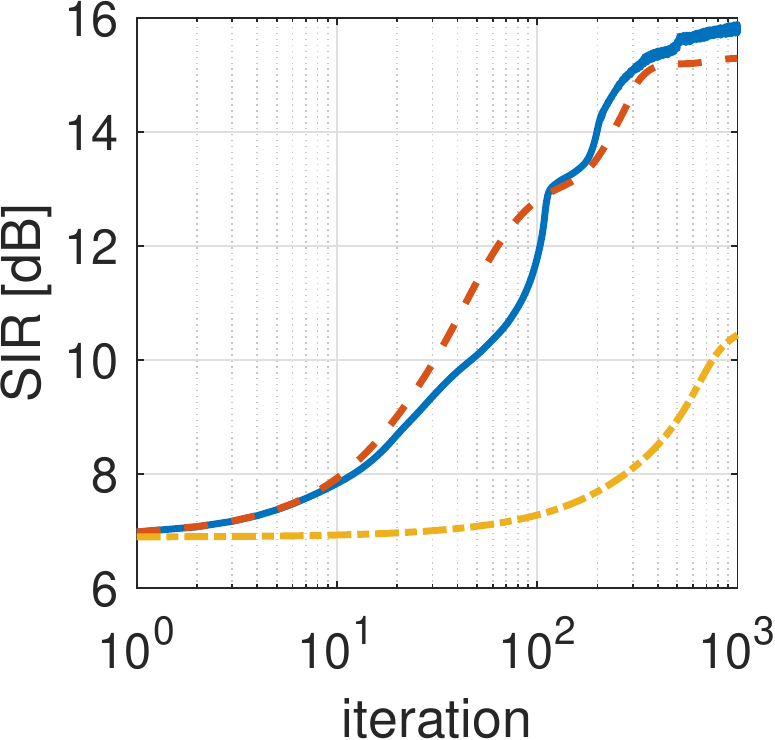}
    \includegraphics[width=0.49\linewidth]{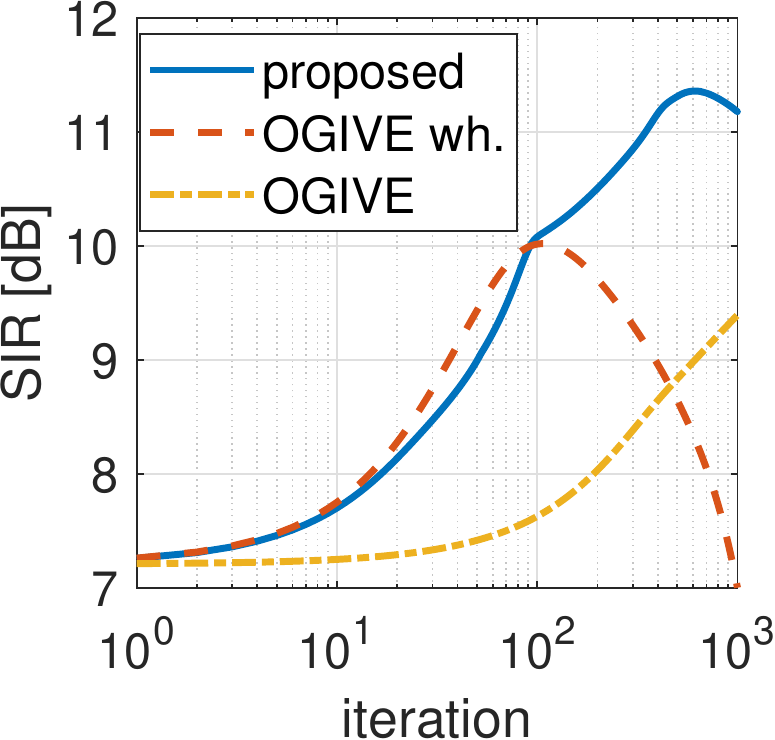}
    \caption{Convergence in terms of SIR of the proposed algorithm and of OGIVE$_{\bf w}$ from \cite{koldovsky2019TSP} when $T_{60}$ is $100$ms (left) and $200$ms (right).}
    \label{fig:fig1}
\end{figure}

\begin{figure}[ht]
    \centering
    \includegraphics[width=0.49\linewidth]{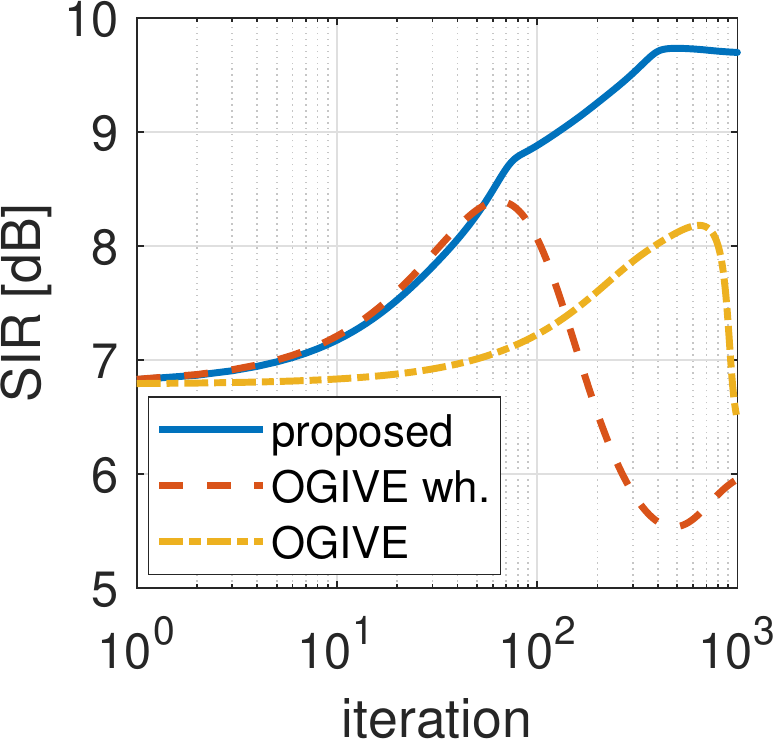}
    \caption{Convergence in terms of SIR as a function of the iteration index when $T_{60}$ is $300$ms.}
    \label{fig:fig2}
\end{figure}

The output SIR achieved by the algorithms is shown in Fig.~\ref{fig:fig1} as a function of the iteration index. The proposed algorithm shows comparable convergence speed like OGIVE$_{\bf w}$ with whitening; the convergence of OGIVE$_{\bf w}$ without whitening is significantly slower. The final SIR achieved by the proposed algorithm is comparable to OGIVE$_{\bf w}$ for $T_{60}=100$ms. For $T_{60}=200$ms, the performance by OGIVE$_{\bf w}$ with wh. decreases after $100$ iterations, possibly due to convergence to a local extreme caused by the permutation problem. A similar phenomenon is observed for $T_{60}=300$ms in Fig.~\ref{fig:fig2}.

\subsection{Dense microphone array}
\begin{figure}[h]
    \centering
    \includegraphics[width=0.95\columnwidth]{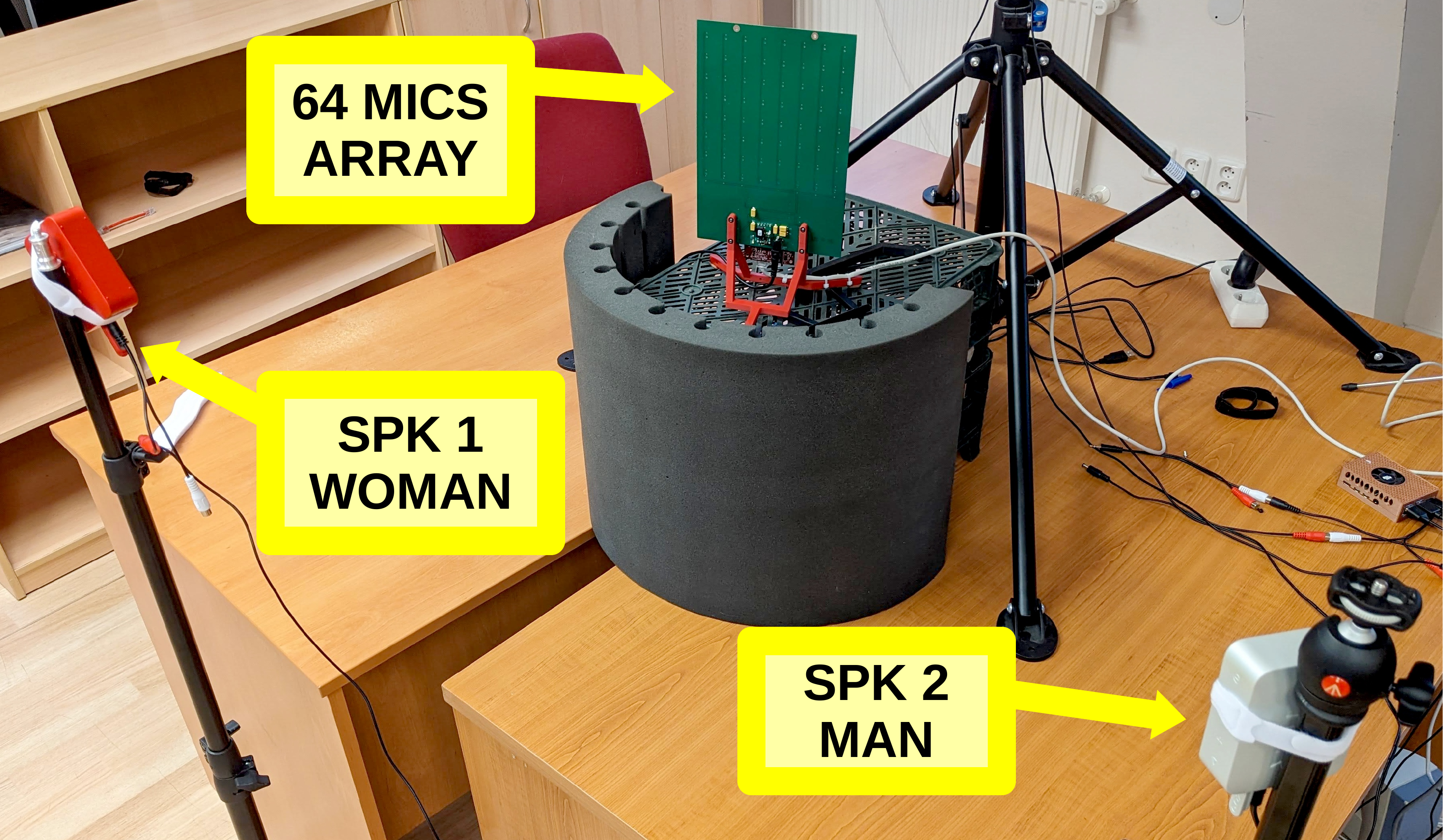}
    \caption{Photo of the experimental setup.}
    \label{fig:setup_photo}
\end{figure}

Here, we describe the results of a real-world experiment. The experimental setup is in an  open-space office, an attic room with dimensions $13\times 8\times 2.5$~m and $T_{60}\approx 500$~ms. Loudspeakers emitting female and male voices are placed at $\approx 0.7$~m from a microphone array at angular positions $40^\circ$ and $-40^\circ$, respectively; a photo of the setup is in Fig.\ref{fig:setup_photo}. The speakers are simulated by JBL GO 2 loudspeakers. 

The microphone array is a dense planar array containing 64 MEMS microphones arranged in the $8\times 8$ regular grid with the horizontal and vertical spacing of $20$~mm. An embedded FPGA controller sends all synchronized sound data observed on microphones through the network, which are then saved and processed in Matlab. The $3.75$~s long recordings were gathered at the $48$~kHz sampling frequency and downsampled to $16$~kHz. In the experiment, we use the first and last column of microphones, i.e., $1$ through $8$ and $57$ through $64$.  

Table~\ref{tab:my_label} shows SIR achieved by the compared algorithms after $200$ iterations when extracting the female speaker; the initial SIR is about $0.5$~dB. The experiment is repeated for various DFT lengths; the shift is always one-quarter of the length of the DFT. The proposed algorithm and OGIVE with whitening achieve their optimum performance when the DFT length is $128$; OGIVE without whitening achieves best for the length of $256$, however, this might be caused by the limited number of iterations. The results suggest that with a dense arrangement of a higher number of  microphones, it is worth choosing a medium length of separating filters.

\begin{table}[h]
    \centering
    \caption{Signal-to-Interference Ratio [dB] achieved in the experiment with dense microphone array}
    \begin{tabular}{c||c|c|c}
 DFT length   & proposed & OGIVE wh. & OGIVE \\
 \hline
 32  & 6.17 &   0.40&   -1.17 \\   
 64  & 9.11 &   2.88&   -0.70 \\
 128 &  9.51&   7.22&   4.23 \\
 256 &  6.28&   6.23&   5.55 \\
 512 &  5.68&   3.88&   3.72 \\
 1024&  3.55&   3.51&   1.72 \\
    \end{tabular}
    
    \label{tab:my_label}
\end{table}

\section{Conclusions}
We have proposed a modified IVE algorithm that operates on the manifold of half-length separating filters. The algorithm does not require any matrix inversion operation and yet has comparably fast convergence as a gradient algorithm using data pre-whitening. However, there is much room for further development and experimental validation. It will be interesting to implement this idea, for example, also for filters of quarter, eighth, or sixteenth length, and, in particular, in a zero-padded STFT domain. The equation \eqref{eq:keyrelation} has an interesting interpretation, namely that the change of any  parametric vector ${\bf v}^\ell$ has a direct effect on the change of all parametric vectors ${\bf w}^k$. This may have interesting implications for the solution of the permutation problem because it means that the algorithm's convergence to a given speaker at a selected frequency significantly affects the gradient at the other frequencies, more so than in the case of conventional frequency-domain IVE.

\balance
\bibliographystyle{hieeetr}

\end{document}